\pdfoutput=1
\documentclass[11pt,twoside,a4paper,cmspaper,final,collab]{cms-tdr}
\def\svnVersion{9a7a1d9}\def\svnDate{2024/10/18}\def\cmsCernNoTag{CERN-EP-2024-107}\def\cmsCernDate{\today}\def\cmsMessage{Published in Physical Review Letters as \href{http://dx.doi.org/10.1103/PhysRevLett.133.161805}{\doi{10.1103/PhysRevLett.133.161805}.}}
\begin{document}\cmsNoteHeader{SMP-22-016}

\newcommand{\fourmuons}{\ensuremath{4\PGm}\xspace}
\newcommand{\ttmmlong}{\ensuremath{\PGt\PGt\PGm\PGm}\xspace}
\newcommand{\Ztautaumumu}{\ensuremath{\PZ\to\ttmmlong}\xspace}
\newcommand{\lllplp}{\ensuremath{\Pell\Pell\Pell'\Pell'}\xspace}
\newcommand{\Zmumumumu}{\ensuremath{\PZ\to\fourmuons}\xspace}
\newcommand{\fourmumass}{\ensuremath{m_{4\PGm}}\xspace}
\newcommand{\pp}{\ensuremath{\Pp\Pp}\xspace}
\newcommand{\RatioSymbol}{\ensuremath{\mathcal{R}_{\ttmmlong}}\xspace}
\newcommand{\RatioSymbolSM}{\ensuremath{\mathcal{R}_{\ttmmlong}^{\text{SM}}}\xspace}
\newcommand{\cllmmtt}{\ensuremath{\mathcal{C}_{\text{LL}}^{2233}}\xspace}
\newcommand{\cllmttm}{\ensuremath{\mathcal{C}_{\text{LL}}^{2332}}\xspace}
\newcommand{\clemmtt}{\ensuremath{\mathcal{C}_{\text{LR}}^{2233}}\xspace}
\newcommand{\clettmm}{\ensuremath{\mathcal{C}_{\text{LR}}^{3322}}\xspace}
\newcommand{\clemttm}{\ensuremath{\mathcal{C}_{\text{LR}}^{2332}}\xspace}
\newcommand{\ceemmtt}{\ensuremath{\mathcal{C}_{\text{RR}}^{2233}}\xspace}
\newcommand{\wc}{\ensuremath{\mathcal{C}}\xspace}
\newcommand{\ci}{\ensuremath{\wc_i}\xspace}
\newcommand{\osixi}{\ensuremath{\mathcal{O}^6_i}\xspace}
\newcommand{\lsm}{\ensuremath{\mathcal{L}_{\text{SM}}}\xspace}
\newcommand{\lsix}{\ensuremath{\mathcal{L}_6}\xspace}
\newcommand{\dr}{\ensuremath{\Delta R}\xspace}
\newcommand{\nzttmm}{\ensuremath{N(\Ztautaumumu)}\xspace}
\newcommand{\nzmmmm}{\ensuremath{N(\Zmumumumu)}\xspace}
\newcommand{\tmvv}{\ensuremath{\PGt\to\PGm\PGn\PGn}\xspace}
\newcommand{\btmvv}{\ensuremath{\mathcal{B}(\tmvv)}\xspace}
\newcommand{\ft}{\ensuremath{f_\PGt}\xspace}
\newcommand{\fm}{\ensuremath{f_\PGm}\xspace}
\newcommand{\aet}{\ensuremath{(A\epsilon)_{\Zmumumumu}}\xspace}
\newcommand{\aem}{\ensuremath{(A\epsilon)_{\Ztautaumumu}}\xspace}
\newcommand{\cdivlsq}{\ensuremath{\mathcal{C}/\Lambda^2}\xspace}
\ifthenelse{\boolean{cms@external}}{\providecommand{\NA}{\ensuremath{\cdots}\xspace}}{\providecommand{\NA}{\ensuremath{\text{---}}\xspace}}
\ifthenelse{\boolean{cms@external}}{\providecommand{\cmsLeft}{upper\xspace}}{\providecommand{\cmsLeft}{left\xspace}}
\ifthenelse{\boolean{cms@external}}{\providecommand{\cmsRight}{lower\xspace}}{\providecommand{\cmsRight}{right\xspace}}

\cmsNoteHeader{SMP-22-016}
\title{Search for the \texorpdfstring{\PZ}{Z} boson decay to \texorpdfstring{\ttmmlong}{tautaumumu} in proton-proton collisions at \texorpdfstring{$\sqrt{s}=13\TeV$}{sqrt(s) = 13 TeV}}

\author*[uzh]{Ben Kilminster}
\author*[uzh]{Stefanos Leontsinis}
\author*[uzh]{Fanqiang Meng}
\author*[uzh]{Arne Reimers}
\author*[uzh]{Yuta Takahashi}

\date{\today}

\abstract{The first search for the \PZ boson decay to \ttmmlong at the CERN LHC is presented, based on data collected by the CMS experiment at the LHC in proton-proton collisions at a center-of-mass energy of 13\TeV and corresponding to an integrated luminosity of 138\fbinv. The data are compatible with the predicted background. For the first time, an upper limit at the 95\% confidence level of 6.9 times the standard model expectation is placed on the ratio of the \Ztautaumumu to \Zmumumumu branching fractions. Limits are also placed on the six flavor-conserving four-lepton effective-field-theory operators involving two muons and two tau leptons, for the first time testing all such operators.}

\hypersetup{
pdfauthor={CMS Collaboration},
pdftitle={Search for the Z boson decay to 2 tau 2 mu in proton-proton collisions at sqrt(s)=13TeV},
pdfsubject={CMS},
pdfkeywords={CMS, Z bosons, muons}}

\maketitle

\begin{figure}[htbp!]
\centering
    \includegraphics[width=0.4\textwidth]{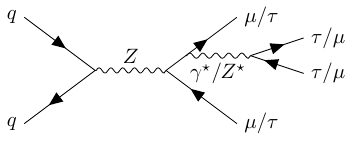}
    \caption{Leading-order Feynman diagram of the decay \Ztautaumumu.}
    \label{fig:diagram}
\end{figure}

\tolerance=800 The large data set collected by the CMS experiment~\cite{CMS:2008xjf} at the CERN LHC facilitates searches for rare processes within the standard model (SM) of particle physics.
The CMS experiment was the first to observe the rare decays $\PZ\to\PJGy\Pell\Pell$~\cite{CMS:2018vcg} and $\PZ\to\lllplp$~\cite{CMS:2012bw} in proton-proton~(\pp) collisions, where \Pell denotes a charged lepton and $\Pell'$ a charged lepton of a possibly different flavor.
At leading order (LO), the latter process occurs via a $\PZ\to\Pell\Pell\to\lllplp$ transition. These transitions are of particular interest because they can receive contributions from hypothetical new particles, such as a \PZpr boson, that can modify the branching fractions predicted by the SM. Theories predicting such new particles have been reported~\cite{Baek:2001kca, Harigaya:2013twa, Altmannshofer:2014cfa, Bell:2014tta, Crivellin:2015mga, Elahi:2015vzh, Altmannshofer:2016oaq, Elahi:2017ppe, Iguro:2020qbc, Buras:2021btx} and exclusive couplings to muons and tau leptons are predicted in many models, making \Ztautaumumu decays, shown in Fig.~\ref{fig:diagram}, important to explore. \par

To derive constraints on possible beyond-the-SM (BSM) effects, the framework of the SM effective field theory~\cite{Brivio:2017vri, Isidori:2023pyp} (SMEFT) is used in this Letter. Here, the SM Lagrangian density \lsm is extended by an infinite series of operators of even canonical dimension six and above, $\mathcal{L} = \lsm + \lsix + \dots$, where $\lsix = 1/\Lambda^2\sum_i\ci\osixi$. 
In this expression, $\Lambda$ is the characteristic energy scale of possible BSM interactions, taken to be $\Lambda=1\TeV$, \osixi are the dimension-six SMEFT operators, and \ci are the corresponding dimensionless Wilson coefficients (WCs). We exclusively consider four-lepton operators and include only the lowest-order contributions from the dimension-six operators to the four-lepton decays of the \PZ boson. We neglect odd-dimensional operators in this Letter because they would induce lepton- and baryon-number violating processes. Stringent constraints on four-muon operators have been derived from the branching fraction of the \Zmumumumu decay~\cite{Boughezal:2020klp}. Throughout this Letter, the notation \Zmumumumu refers to the prompt decay of the \PZ boson to four muons that do not originate from decays of tau leptons or hadrons. However, four-lepton operators of dimension six involving two tau leptons and two muons are currently poorly constrained because of a lack of measurements of processes to which they could contribute at LO. Only one of these operators is constrained by measurements of the tau lepton decay to muons and neutrinos via charged-current interactions, whereas all others have never been probed~\cite{Falkowski:2015krw, Boughezal:2020klp}.

\tolerance=800 In this Letter, we report the first LHC search for the \PZ boson decay to two tau leptons and two muons relative to the \PZ boson decay to four muons. This search benefits from a partial cancellation of experimental systematic uncertainties that affect both decay modes. The two tau leptons are reconstructed via their decays to muons and neutrinos, hence the \Ztautaumumu decay is studied in the final state with four muons with a total electric charge of zero. In this analysis, we use \pp collision data recorded by the CMS experiment in the years 2016--2018 at a center-of-mass energy of 13\TeV that correspond to an integrated luminosity of 138\fbinv~\cite{CMS-LUM-17-003, CMS-PAS-LUM-17-004, CMS-PAS-LUM-18-002}. Measurements of four-fermion events possibly produced by the \PZ boson decay studied here have previously been reported at LEP~\cite{ALEPH:1991oac,OPAL:1992qxl,DELPHI:1993kec,L3:1993wwa}. Tabulated results are provided in the HEPData record for this analysis~\cite{hepdata}.\par

The CMS apparatus~\cite{CMS:2008xjf} is a multipurpose, nearly hermetic detector, designed to trigger on~\cite{CMS:2020cmk, CMS:2016ngn} and identify~\cite{CMS:2020uim,CMS-DP-2020-021,CMS:2018rym,CMS:2014pgm} electrons, muons, photons, and hadrons. A global particle-flow algorithm~\cite{CMS:2017yfk} aims to reconstruct all individual particles in an event, combining information provided by the all-silicon tracking detector and by the crystal electromagnetic and brass-scintillator hadron calorimeters, operating inside a 3.8\unit{T} superconducting solenoid, with data from the gas-ionization muon detectors embedded in the flux-return yoke outside the solenoid. The reconstructed particles are used to identify \PGt leptons and jets and to measure the missing transverse momentum~\cite{CMS:2018jrd, CMS:2016lmd, CMS:2019ctu}.

\tolerance=800 Events of interest are selected using a two-tiered trigger system. The first level, composed of custom hardware processors, uses information from the calorimeters and muon detectors to select events at a rate of around 100\unit{kHz} within a fixed latency of about 4\mus~\cite{CMS:2020cmk}. The second level, known as the high-level trigger, consists of a farm of processors running a version of the full event reconstruction software optimized for fast processing and reduces the event rate to around 1\unit{kHz} before data storage~\cite{CMS:2016ngn}. In this analysis, a single-muon trigger is used to record events with at least one isolated muon candidate having transverse momentum ${\pt>27~(24)\GeV}$ in the 2017 (2016 and 2018) data-taking period(s).\par

Muons are measured in the pseudorapidity range $\abs{\eta}<2.4$, with detection planes made using three technologies: drift tubes, cathode strip chambers, and resistive-plate chambers. The efficiency to reconstruct and identify muons is greater than 96\% over the full $\eta$ range. Matching muons to tracks measured in the silicon tracker results in a relative \pt resolution, for muons with \pt up to 100\GeV, of 1\% in the barrel and 3\% in the endcaps~\cite{CMS:2018rym}. The primary vertex of an event is taken to be the vertex corresponding to the most energetic scattering, evaluated using tracking information alone, as described in Ref.~\cite{CMS-TDR-15-02}.

\tolerance=800 In this analysis, all reconstructed muons are required to pass a set of loose identification criteria, although the highest-\pt muon must satisfy a tighter condition. Both the loose and the tight criteria, are defined in Ref.~\cite{CMS:2018rym}. All muons are required to be isolated from other activity in the event, quantified by the scalar \pt sum of hadrons and photons with an angular distance from the muon $\dr=\sqrt{\smash[b]{(\Delta\eta)^2+(\Delta\phi)^2}}<0.4$, where $\phi$ is the azimuthal angle. This sum does not include contributions from charged particles identified as originating from pileup vertices and is corrected for the remaining average hadronic activity in an event due to pileup. The ratio of this corrected sum to the muon \pt is required to be smaller than 0.15 for the muon with the highest \pt and at most 0.25 for all other muons.\par

Processes leading to a final state with four prompt charged leptons via quark-initiated single or double electroweak gauge boson production are simulated at next-to-LO (NLO) in quantum chromodynamics, including all lepton flavors, using the \POWHEG~v2~\cite{Nason:2004rx,Frixione:2007vw,Alioli:2010xd,Melia:2011tj,Nason:2013ydw} Monte Carlo event generator. The invariant mass of any two generated same-flavor leptons with opposite electric charges is required to exceed 4\GeV. Events containing the lepton flavor combinations targeted in this analysis, \ttmmlong and \fourmuons, are isolated from this sample using generator-level information. 

Events of triple vector boson ({\PV}{\PV}{\PV}) production and top quark-antiquark ({\ttbar}) production in association with a \PZ boson are generated at NLO with \MGvATNLO~v2.6.5~\cite{Alwall:2014hca}. Higgs boson (H) production and its subsequent decay to four charged leptons is simulated at NLO with \POWHEG~v2~\cite{Nason:2009ai,Bagnaschi:2011tu} and \textsc{jhugen} v.7.0.11~\cite{Gao:2010qx,Bolognesi:2012mm,Anderson:2013afp}. Small background contributions due to combinations of prompt and nonprompt charged leptons come from \ttbar and quark-initiated double vector boson ({\PV}{\PV}) production. The former process, as well as {\PW}{\PW} and {\PZ}{\PZ} production with decays to two charged leptons and two neutrinos, are simulated at NLO with \POWHEG~v2~\cite{Frixione:2007nw, Melia:2011tj, Nason:2013ydw}. All other {\PV}{\PV} processes producing two or three charged leptons are generated at NLO using \MGvATNLO~v2.6.5. Gluon-induced {\PV}{\PV} production is negligible compared with the corresponding quark-initiated process and not included in the simulated samples.

All events are generated using the NNPDF3.1 next-to-NLO parton distribution functions~\cite{NNPDF:2017mvq}. The parton shower and subsequent hadronization are simulated with \PYTHIA~8.205~\cite{Sjostrand:2014zea} and the underlying event is modeled using the CP5 tune~\cite{CMS:2019csb}. Additional inelastic \pp interactions in the same or adjacent bunch crossings (pileup) are simulated for all processes, and events are reweighted to match the measured number of pileup interactions in the data. The full CMS detector is simulated using \GEANTfour~\cite{Agostinelli:2002hh}.

The generator-level phase space is defined by a set of criteria imposed on events at the generator level. The presence of exactly four muons that form two pairs with opposite electric charges consistent with the decays \Zmumumumu and \Ztautaumumu, the latter followed by two \tmvv decays, is required. The invariant mass of the four muons, \fourmumass, must satisfy $40<\fourmumass<100\GeV$. The pair of oppositely charged muons with the highest invariant mass is required to have $12<m_{\PGm\PGm}^{\text{max}}<75\GeV$, whereas for all other such pairs $m_{\PGm\PGm}>4\GeV$ is imposed.

The same requirements are applied to both data and simulated events at the reconstructed level as well. Additionally, the \pt-leading muon must have $\pt>29~(26)\GeV$ in the 2017 (2016 and 2018) data-taking period(s), reflecting the single-muon trigger threshold. Minimum transverse momenta of 3.5\GeV are required for the two muons with the second- and third-highest \pt in the event. The fourth muon must satisfy $\pt>3.5\GeV$ for $\abs{\eta}<1.2$ and $\pt>2.5\GeV$ for $1.2<\abs{\eta}<2.4$ to ensure it reaches the outer muon detectors, considering the CMS magnetic field and the energy loss of the muon traversing the detector. Finally, each muon is required to have an angular distance of $\dr>0.02$ from any other muons, and the four selected muons must form a valid vertex.

The background due to nonprompt muons, most copiously produced in quantum chromodynamics multijet events via decays of bottom or charm quarks, is estimated using control samples in the data. Three additional, signal-depleted event categories are defined by inverting the requirement on the isolation of at least one of the three \pt-subleading muon candidates, on their total electric charge, or both. The inverted variables are independent, and the shape of the four-muon invariant mass distribution is consistent between any of these control regions and the search region. The shape and normalization of the background due to nonprompt muons in the latter is obtained by scaling the \fourmumass distribution of data in the control region with inverted isolation criterion, from which the residual contribution of prompt muons is subtracted using simulated events. The scale factor is derived from the same extrapolation in the regions with inverted charge requirement. A correction accounting for differences between the regions with and without inverting the total electric charge criterion is derived by repeating the extrapolation between two subsets of events with inverted muon isolation. 

A 47\% uncertainty in the extrapolation factor used in the estimation of the background from nonprompt muons is due to the statistical uncertainty in the data in the signal-depleted regions; it is the dominant uncertainty in this search. Additional large sources of uncertainty are the number of events in the measurements and the simulation as well as the estimate of the nonprompt muon background in the search region. Minor experimental uncertainties in the efficiencies of different criteria imposed on the muons affect the shape and normalization of all simulated processes. The minor effect (${<}10\%$) of theoretical uncertainties in the renormalization and factorization scales, parton distribution functions, and strong coupling $\alpS(m_\PZ)$ on the \fourmumass distribution are also included. For the former, the change in acceptance is estimated from independent variations of each scale by a factor of 0.5 or 2, whereas the uncertainty due to the latter two is estimated following the description in Ref.~\cite{Butterworth:2015oua}. The integrated luminosities for the 2016, 2017, and 2018 data-taking years have 1.2--2.5\% individual uncertainties~\cite{CMS-LUM-17-003,CMS-PAS-LUM-17-004,CMS-PAS-LUM-18-002} that affect only the rate of simulated processes, and the overall luminosity uncertainty for the 2016--2018 period is 1.6\%. The uncertainties in the SM cross section of the individual simulated backgrounds amount to 5--25\% depending on the process. The effect of all systematic uncertainties is also propagated to the background with nonprompt muons estimated from data via the subtraction of simulated events with prompt muons. 

The yield of \Ztautaumumu events is extracted from a binned maximum likelihood template fit of the expected signal and background to the data in the \fourmumass distribution, which is performed with the CMS statistical analysis tool \textsc{Combine}~\cite{CAT-23-001}. Because of the neutrinos arising from the tau lepton decays, \Ztautaumumu events are expected to form a broad distribution in \fourmumass in the region below the \PZ boson mass. Each systematic uncertainty is taken into account as a nuisance parameter in this fit. Uncertainties not affecting the shape of the distribution are modeled with log-normal probability distributions, whereas those affecting both the shape and normalization are instead modeled with Gaussian distributions.

\tolerance=1500 The number of \Zmumumumu events is scaled by an unconstrained parameter in the fit. A second unconstrained parameter, $r$, is used to scale the ratio of the number of \Ztautaumumu and \Zmumumumu events. Both parameters are defined with respect to the SM expectation, which corresponds to parameter values of 1. In the generator-level phase space defined by the requirements on the dimuon and four-muon invariant mass discussed above, the ratio \RatioSymbol of the \Ztautaumumu to \Zmumumumu branching fractions is given by 
\begin{equation}
    \RatioSymbol = \frac{\nzttmm}{\nzmmmm} \frac{\aet}{\aem}\frac{1}{\mathcal{B}^2({\tmvv})}\frac{\ft}{\fm}.
    \label{eq:formula}
\end{equation}
\tolerance=1500 Here, \nzttmm and \nzmmmm are the event yields of the respective processes, where only \tmvv decays are considered for the former. The acceptance and efficiency of selecting events from the generator-level region are denoted by $A$ and $\epsilon$, respectively. Their products are $\aet=(6.40\pm0.01)\% $ and ${\aem=(1.32\pm0.02)\%}$ for the respective \PZ boson decay modes and are determined from simulation. The parameters \ft and \fm are the fractions of selected events in the simulated signal and reference sample, respectively, that are due to the decay of a single \PZ boson produced in the $s$ channel. The remaining events are due to {\PV}{\PV} production leading to the same final state. The parameters, which are computed by \MGvATNLO, are $\ft = (83.9 \pm 0.1)\%$ and $\fm = (85.5 \pm 0.2)\%$, where the uncertainties are statistical. The known branching fraction of the tau lepton decay to a muon and the corresponding neutrinos is ${\btmvv=(17.39\pm 0.04)\%}$~\cite{ParticleDataGroup:2022pth}. The SM expectation in the generator-level phase space, derived using generator-level information of events simulated with \POWHEG~v2 at NLO, is $\RatioSymbolSM=0.90\pm0.02$, where the uncertainty includes statistical uncertainties in the number of generated events of each process and the uncertainties in the other factors of Eq.~\eqref{eq:formula}, as stated above.\par

\begin{figure}[th]
\centering
    \includegraphics[width=0.49\textwidth]{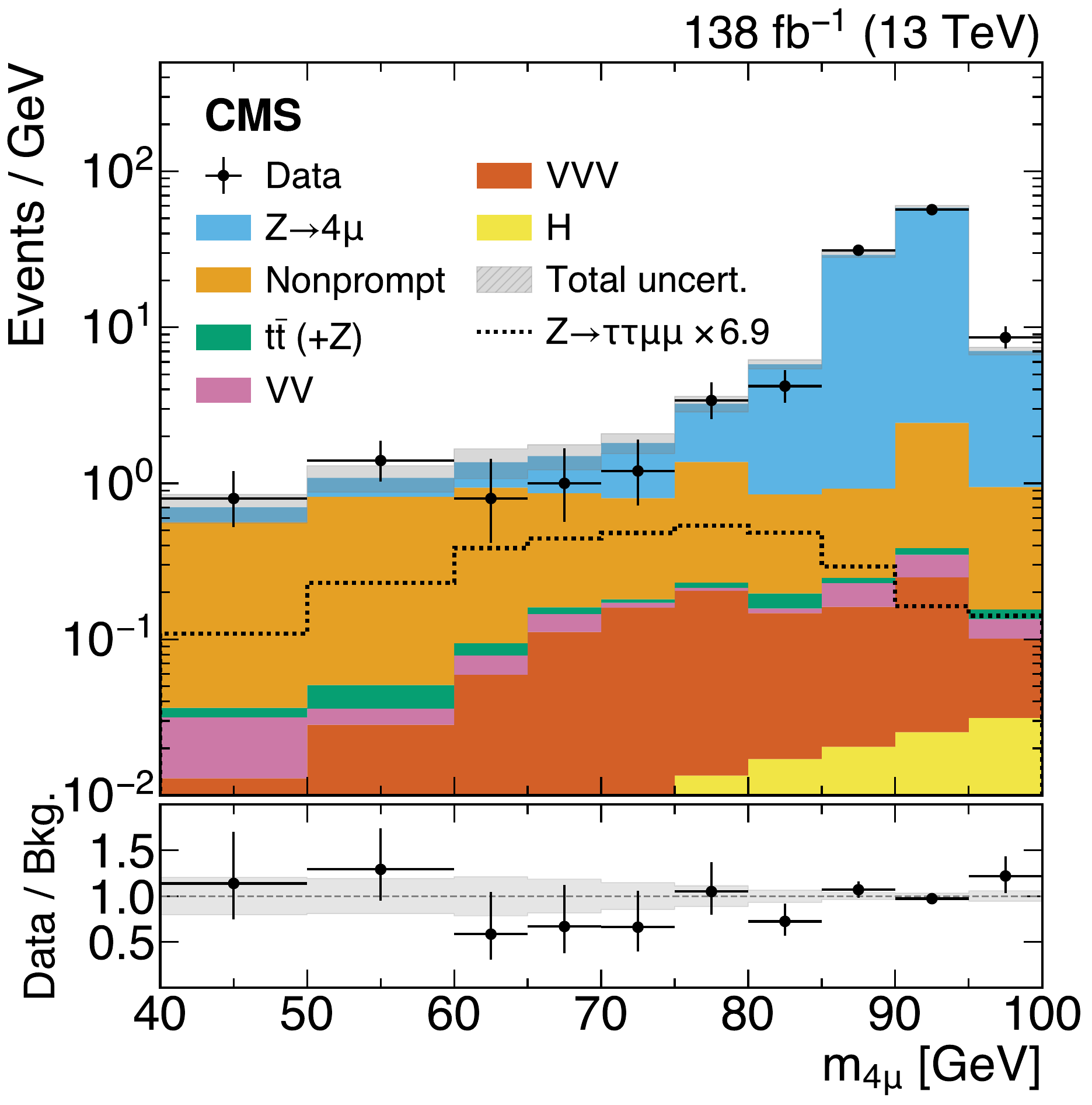}
    \caption{Distribution of \fourmumass after the maximum likelihood fit of the background-only model (stacked histograms) to the data (black points). The nuisance parameters are set to their post-fit values and the signal (black dotted line) is overlaid, scaled to the upper limit on its cross section of 6.9 times the SM expectation. The gray shaded areas in both panels correspond to the total uncertainty in the background prediction. The black vertical bars indicate the statistical uncertainty in the data.}
    \label{fig:mass}
\end{figure}

In the combined fit of signal and background to the data, the best fit ratio of branching fractions is extracted from the fitted value of $r$, 
\begin{equation}
\RatioSymbol = r\RatioSymbolSM = -12.5^{+6.8}_{-8.2}.
\end{equation}
This agrees with the SM expectation within two standard deviations and has a negative value due to the deficit of data in the region $60<\fourmumass<75\GeV$. 

A second maximum likelihood fit of the background-only expectation to the data is performed. The \fourmumass distribution after this fit is shown in Fig.~\ref{fig:mass}. The observed data agree with the background prediction within the uncertainties. The postfit number of \Zmumumumu events is consistent between the signal-and-background and background-only fits with values of $1.10\pm0.07$ and $1.09\pm0.07$ times the SM expectation, respectively.

In addition, upper limits at the 95\% confidence level ({\CL}) are placed on \RatioSymbol for the first time, derived using the \CLs technique~\cite{Junk:1999kv, Read_2002} and the \textsc{Combine} tool~\cite{CAT-23-001}. The observed and median expected upper limits on \RatioSymbol are 6.2 and 10.0, respectively, reflecting the deficit of data discussed above. The intervals $\left[7.3,13.9\right]$ ($\left[5.5,18.8\right]$) include 68 (95\%) of the distribution of expected limits under the background-only hypothesis. The observed and median expected limits correspond to 6.9 and 11.1 times the SM expectation of $\RatioSymbolSM = 0.90$, respectively.

The 95\% \CL upper limits are used in the SMEFT framework to derive constraints on potential BSM contributions to \RatioSymbol as suggested in Ref.~\cite{Boughezal:2020klp}. In this Letter, the notation $\wc_{\text{AB}}^{ijkl}$ is used for the WC corresponding to the four-lepton operator $(\overline{\Pell}^i_{\text{A}}\gamma_{\mu}\Pell^j_{\text{A}})(\overline{\Pell}^k_{\text{B}}\gamma^{\mu}\Pell^l_{\text{B}})$ contributing to the process $\Pell^l\to\Pell^k\Pell^j\Pell^i$, where the superscripts denote the lepton generation and A and B indicate the chirality. Constraints on the WCs are derived in terms of $\wc/\Lambda^2$ with $\Lambda = 1\TeV$.

In this analysis, all six SMEFT operators in the so-called Warsaw basis~\cite{Grzadkowski:2010es} that have a canonical dimension of six and involve two muons and two tau leptons are considered. These operators affect only the \Ztautaumumu but not the \Zmumumumu decay. The effect of each operator on Eq.~\eqref{eq:formula}, relative to the SM prediction, is computed using \MGvATNLO~v2.7.2 with the \textsc{SMEFTsim}~3.0 package~\cite{Brivio:2017btx, Brivio:2020onw}. The multiplicative BSM correction to each factor in Eq.~\eqref{eq:formula} is expressed in terms linear or quadratic in the corresponding WC, where the linear terms are due to interference of the BSM process with the SM and the quadratic terms correspond to pure BSM contributions. The total correction to \RatioSymbolSM is given by the product of the individual corrections to each factor. In what follows, distinctions between linear and quadratic terms refer to the order of the individual correction factors, their product therefore also depends on higher orders of the WCs in both cases.

\begin{figure}[th]
    \centering
    \includegraphics[width=0.49\textwidth]{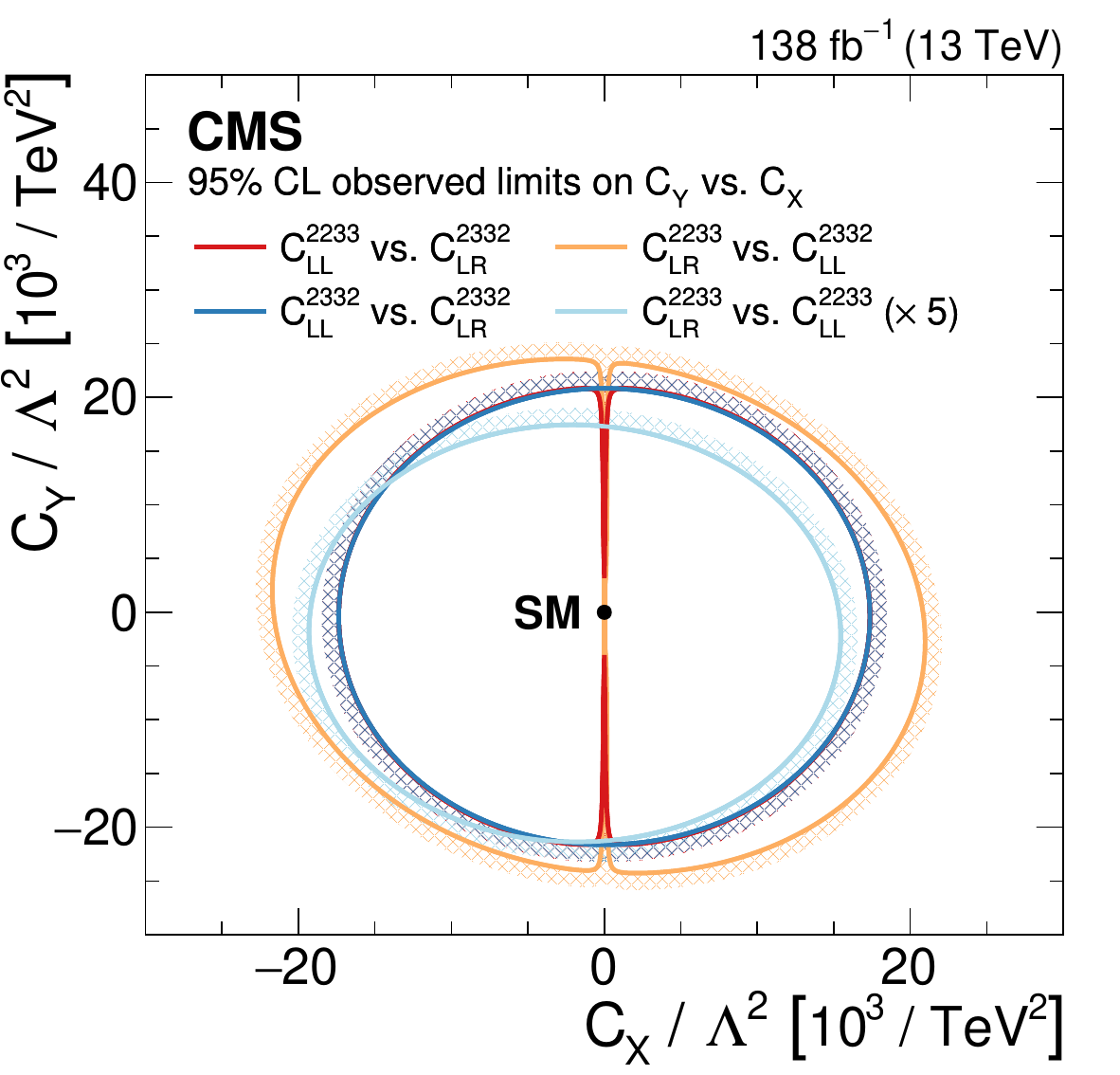}
    \includegraphics[width=0.49\textwidth]{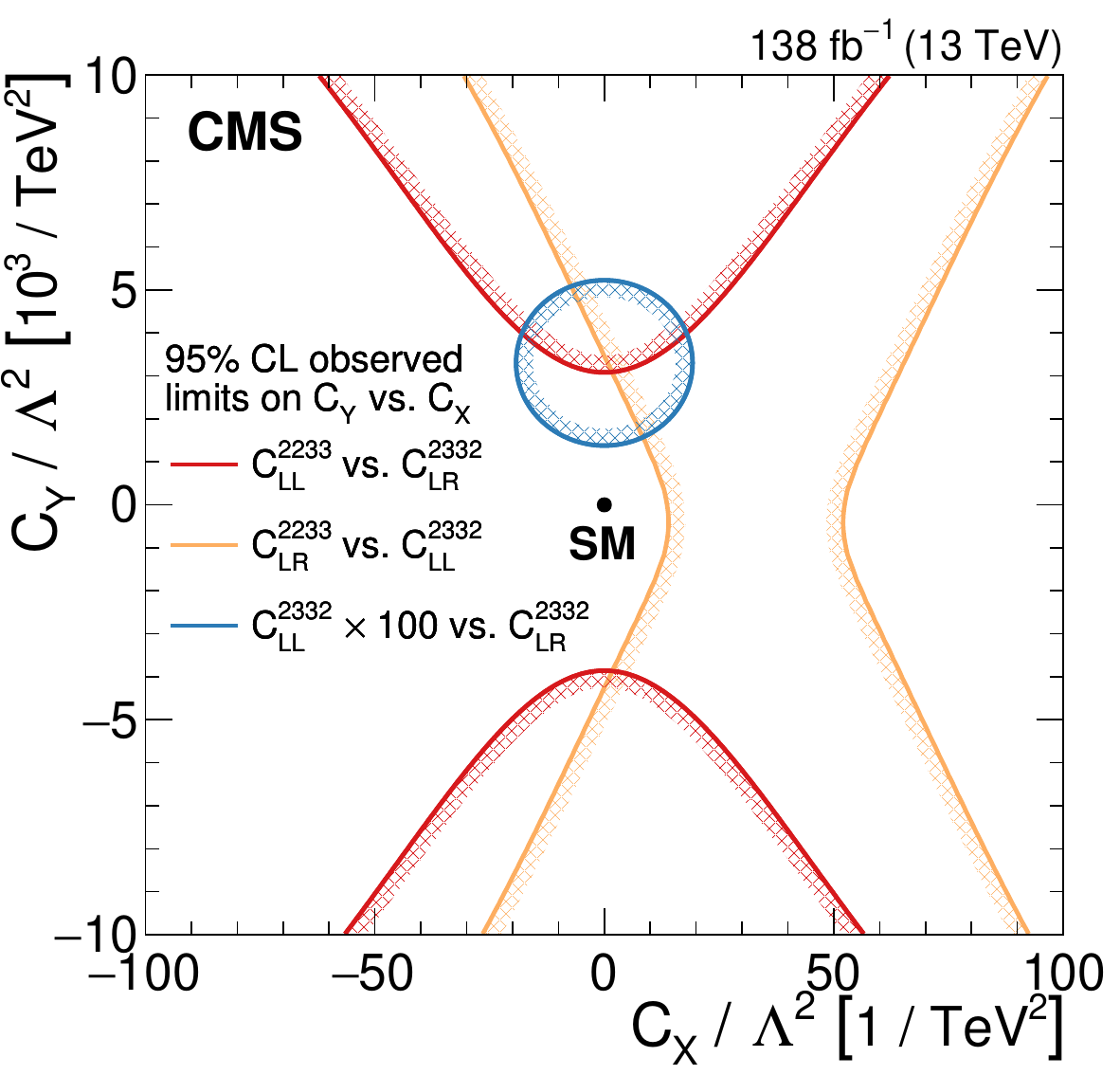}
    \caption{Observed limits at the 95\% \CL on different combinations of two Wilson coefficients (colors) showing the full ranges (\cmsLeft) and an enlarged view of the region around values of 0 (\cmsRight). In the \cmsLeft (\cmsRight) figure, the limit on \clemmtt vs. \cllmmtt (only {\cllmttm}) is scaled by a factor of 5 (100) for visibility. The SM expectation is indicated by a black dot. The hatches indicate the excluded side of each limit.}
    \label{fig:WCLimits}
\end{figure}

When we consider one nonzero WC at a time, the intervals allowed at the 95\% \CL for \cdivlsq are given in Table~\ref{tab:1dlimits}. When we consider two nonzero WCs and linear and quadratic terms, the regions allowed and excluded at the 95\% \CL are shown in Fig.~\ref{fig:WCLimits} for four representative combinations of WCs that exhibit characteristic features discussed below.
These are the first constraints on all flavor-conserving four-lepton WCs involving two muons and two tau leptons with the exception of \cllmttm~\cite{Falkowski:2015krw, Boughezal:2020klp}, which was limited more stringently by Ref.~\cite{Falkowski:2015krw}. 

The effect of any WC on the event yield is dominant compared to BSM contributions to the signal acceptance and \ft. It was verified that BSM shape distortions of the \fourmumass distribution after applying the event selection are negligible compared to changes in the overall event yield. The expected value of \btmvv in Eq.~\eqref{eq:formula} is only affected by two WCs, \cllmttm and \clemttm. For these WCs, however, it has the dominant impact on the determination of the allowed and excluded regions and leads to a larger allowed parameter space when considering quadratic terms as shown in Fig.~\ref{fig:WCLimits} (\cmsLeft). The limit on a combination of WCs including neither \cllmttm nor \clemttm (light blue) is approximately 5 times more stringent than the limits on combinations that include one or both of these WCs. For combinations that include only \clemttm (red), which enhances \btmvv, the limit is most stringent for $\clemttm = 0$ and relaxes for nonzero values as shown in Fig.~\ref{fig:WCLimits} (\cmsRight). Including only \cllmttm (orange), which causes negative interference with the SM prediction of \btmvv, leads to an excluded region at small positive values of \cllmttm. Considering both \clemttm and \cllmttm, their combined effect results in a second excluded region shown in Fig.~\ref{fig:WCLimits} (\cmsRight, dark blue).
In this analysis, external constraints on \btmvv are neglected, but produce stronger constraints on \cllmttm~\cite{Falkowski:2015krw} and \clemttm than those presented here due to the small uncertainty in the measured value of \btmvv~\cite{ParticleDataGroup:2022pth}.

\begin{table}[th]
    \topcaption{Intervals allowed at the 95\% \CL for \cdivlsq for all 6 dimension-six Wilson coefficients conserving lepton flavor and involving two muons and two tau leptons. The second column shows the allowed intervals considering only terms linear in \wc, in which case the lower and upper bounds originate from unphysical negative event yields and the limit on \RatioSymbol, respectively. For the third column, quadratic terms are considered as well and all bounds are determined by the limit on \RatioSymbol.}
    \label{tab:1dlimits}
    \centering
    \renewcommand{\arraystretch}{1.1}
    \begin{scotch}{lll}
                                     & \multicolumn{2}{l}{95\% \CL allowed region for \cdivlsq [$10^3/\TeV^2$]} \\
    \wc                               & Only linear terms    & Linear and quadratic terms        \\\hline
    \rule{0pt}{2ex}\cllmmtt                         & [$-2.7$, 15.2]  & [$-3.8$, 3.1]                     \\
    \cllmttm                         & [$-2.74$, 0.01] & [$-21.62$, 0.01] $\cup$ [0.05, 20.82] \\
    \clemmtt                         & [$-3.0$, 18.5]  & [$-4.3$, 3.5]                     \\
    \clettmm                         & [$-3.0$, 17.2]  & [$-4.2$, 3.4]                     \\
    \clemttm                         & \NA             & [$-17.4$, 17.4]                   \\
    \ceemmtt                         & [$-3.6$, 20.0]  & [$-4.4$, 3.6]              
    \end{scotch}
\end{table}

When we consider only linear terms, each factor in Eq.~\eqref{eq:formula} may take on unphysical values at sufficiently large absolute values of the WCs, such as a negative event yield or $\ft\notin [0, 1]$. The allowed regions given in Table~\ref{tab:1dlimits} are obtained excluding these and other unphysical scenarios. Considering only linear terms, all lower (upper) limits originate from negative event yields (the limit on \RatioSymbol). No linear terms exist for \clemttm because the corresponding operator cannot interfere with the SM process. When considering both linear and quadratic terms, no unphysical regimes are reached within the limits derived from the upper limit on \RatioSymbol.

In summary, the first search for the \PZ boson decay to \ttmmlong at the LHC has been presented. The sample of proton-proton collision data analyzed was collected at $\sqrt{s}=13\TeV$ by the CMS experiment and corresponds to an integrated luminosity of 138\fbinv. No excess over the standard model background prediction was observed. For the first time, an upper limit at 95\% confidence level of 6.2 was placed on the ratio of the \Ztautaumumu to \Zmumumumu branching fractions in the phase space considered, equivalent to 6.9 times the standard model expectation of 0.90. Also for the first time, constraints were placed on all flavor-conserving four-lepton Wilson coefficients involving two muons and two tau leptons.

\begin{acknowledgments}

We congratulate our colleagues in the CERN accelerator departments for the excellent performance of the LHC and thank the technical and administrative staffs at CERN and at other CMS institutes for their contributions to the success of the CMS effort. In addition, we gratefully acknowledge the computing centers and personnel of the Worldwide LHC Computing Grid and other centers for delivering so effectively the computing infrastructure essential to our analyses. Finally, we acknowledge the enduring support for the construction and operation of the LHC, the CMS detector, and the supporting computing infrastructure provided by the following funding agencies: SC (Armenia), BMBWF and FWF (Austria); FNRS and FWO (Belgium); CNPq, CAPES, FAPERJ, FAPERGS, and FAPESP (Brazil); MES and BNSF (Bulgaria); CERN; CAS, MoST, and NSFC (China); MINCIENCIAS (Colombia); MSES and CSF (Croatia); RIF (Cyprus); SENESCYT (Ecuador); ERC PRG, RVTT3 and MoER TK202 (Estonia); Academy of Finland, MEC, and HIP (Finland); CEA and CNRS/IN2P3 (France); SRNSF (Georgia); BMBF, DFG, and HGF (Germany); GSRI (Greece); NKFIH (Hungary); DAE and DST (India); IPM (Iran); SFI (Ireland); INFN (Italy); MSIP and NRF (Republic of Korea); MES (Latvia); LMTLT (Lithuania); MOE and UM (Malaysia); BUAP, CINVESTAV, CONACYT, LNS, SEP, and UASLP-FAI (Mexico); MOS (Montenegro); MBIE (New Zealand); PAEC (Pakistan); MES and NSC (Poland); FCT (Portugal); MESTD (Serbia); MCIN/AEI and PCTI (Spain); MOSTR (Sri Lanka); Swiss Funding Agencies (Switzerland); MST (Taipei); MHESI and NSTDA (Thailand); TUBITAK and TENMAK (Turkey); NASU (Ukraine); STFC (United Kingdom); DOE and NSF (USA).
\end{acknowledgments}

\bibliography{auto_generated}
\cleardoublepage \appendix\section{The CMS Collaboration \label{app:collab}}\begin{sloppypar}\hyphenpenalty=5000\widowpenalty=500\clubpenalty=5000
\cmsinstitute{Yerevan Physics Institute, Yerevan, Armenia}
{\tolerance=6000
A.~Hayrapetyan, A.~Tumasyan\cmsAuthorMark{1}\cmsorcid{0009-0000-0684-6742}
\par}
\cmsinstitute{Institut f\"{u}r Hochenergiephysik, Vienna, Austria}
{\tolerance=6000
W.~Adam\cmsorcid{0000-0001-9099-4341}, J.W.~Andrejkovic, T.~Bergauer\cmsorcid{0000-0002-5786-0293}, S.~Chatterjee\cmsorcid{0000-0003-2660-0349}, K.~Damanakis\cmsorcid{0000-0001-5389-2872}, M.~Dragicevic\cmsorcid{0000-0003-1967-6783}, P.S.~Hussain\cmsorcid{0000-0002-4825-5278}, M.~Jeitler\cmsAuthorMark{2}\cmsorcid{0000-0002-5141-9560}, N.~Krammer\cmsorcid{0000-0002-0548-0985}, A.~Li\cmsorcid{0000-0002-4547-116X}, D.~Liko\cmsorcid{0000-0002-3380-473X}, I.~Mikulec\cmsorcid{0000-0003-0385-2746}, J.~Schieck\cmsAuthorMark{2}\cmsorcid{0000-0002-1058-8093}, R.~Sch\"{o}fbeck\cmsorcid{0000-0002-2332-8784}, D.~Schwarz\cmsorcid{0000-0002-3821-7331}, M.~Sonawane\cmsorcid{0000-0003-0510-7010}, S.~Templ\cmsorcid{0000-0003-3137-5692}, W.~Waltenberger\cmsorcid{0000-0002-6215-7228}, C.-E.~Wulz\cmsAuthorMark{2}\cmsorcid{0000-0001-9226-5812}
% ...
\par}
\vskip\cmsinstskip
\dag:~Deceased\\
$^{1}$Also at Yerevan State University, Yerevan, Armenia\\
$^{2}$Also at TU Wien, Vienna, Austria\\
% ...
\end{sloppypar}
\end{document}